\documentclass[prl,aps,twocolumn]{revtex4}
\usepackage{graphicx}
\usepackage{dcolumn}
\usepackage{bm}
\usepackage{amsmath}

\newcommand{\kk}{{\mathbf k}}
\newcommand{\Mc}{{\mathcal M}}
\newcommand{\Gc}{{\mathcal G}}
\newcommand{\vh}{\hat{v}}
\newcommand{\vhd}{\hat{v}^\dagger}
\newcommand{\ah}{\hat{a}}
\newcommand{\ahd}{\hat{a}^\dagger}
\newcommand{\bh}{\hat{b}}
\newcommand{\bhd}{\hat{b}^\dagger}

\begin{document}
\title{Quantum vacuum radiation spectra from a semiconductor microcavity \\ with a time-modulated vacuum Rabi frequency}
\author{Simone \surname{De Liberato}$^{1,2}$}
\author{Cristiano Ciuti$^{1}$}
\affiliation{$^1$Laboratoire Mat\'eriaux et Ph\'enom\`enes
Quantiques, UMR 7162, Universit\'e Paris 7, 75251 Paris, France}
\affiliation{$^2$Laboratoire Pierre Aigrain, UMR 8551, \'Ecole
Normale Sup\'erieure, 24 rue Lhomond, 75005 Paris, France}
\author{Iacopo Carusotto}
\affiliation{CRS BEC-INFM and Dipartimento di Fisica,
Universit\`{a} di Trento, I-38050 Povo, Italy}
\date{\today}

\begin{abstract}
We develop a general theory of the quantum vacuum radiation
generated by an arbitrary time-modulation of the vacuum Rabi
frequency for an intersubband transition of a doped quantum well
system embedded in a semiconductor planar microcavity. Both
non-radiative and radiative losses are included within an
input-output quantum Langevin framework. The intensity and the
main spectral signatures of the extra-cavity emission are
characterized as a function of the amplitude and the frequency of
the vacuum modulation. A significant amount of photon pairs, which
can largely exceed the black-body radiation in the mid and far
infrared, is shown to be produced with realistic parameters. For a
large amplitude resonant modulation, a parametric oscillation
regime can be also achieved.
\end{abstract}

 \maketitle The radiation generated by a time-modulation of
the quantum vacuum is a very general and fascinating phenomenon,
and has been predicted to occur in a variety of physical systems
ranging from non-uniformly accelerated boundaries (dynamical
Casimir effect ~\cite{RMP,dyn_Casimir}) to semiconductors with
rapidly changing dielectric properties~\cite{Yablonovitch}. These
quantum vacuum phenomena have some analogies with the
Unruh-Hawking radiation\cite{Hawking} in the curved space-time
around a black hole. Recent years have seen the appearance of a
number of proposals to enhance the intensity of the quantum vacuum
radiation, exploting, e.g., high-finesse Fabry-P\'erot~
resonators\cite{Reynaud}, a time-modulation of the dielectric
constant of a monolithic solid-state cavity~\cite{epsilon_var}, or
even the reflectivity change of semiconductor mirrors induced by
an ultrafast photogeneration of carriers~\cite{Braggio}. Still,
the very weak intensity of the emitted radiation has so far
hindered its experimental observation.

Planar semiconductor microcavities embedding a doped multiple
quantum well structure have attracted a considerable interest in
the last few years. As demonstrated by several spectroscopic
experiments in the mid infrared
range~\cite{Dini_PRL,Dupont,Aji_APL,SNS_unpub,SST,Luca}, the
strong coupling between the cavity mode and the electronic
transition between the first two quantum well subbands results in
an elementary excitation spectrum consisting of intersubband
polaritons, i.e. linear superpositions of cavity photon and
intersubband excitation states. The most interesting property of
these systems from the point of view of the quantum vacuum
radiation is given by the large vacuum Rabi frequency $\Omega_R$,
being as high as a significant fraction of the intersubband
transition frequency $\omega_{12}$~\cite{Ciuti_vacuum}. In this
unusual ultrastrong coupling regime~\cite{Ciuti_vacuum,Ciuti_PRA},
the antiresonant terms of the vacuum Rabi coupling play in fact a
significant role and the quantum ground state of the system is a
squeezed vacuum state containing a significant amount of
correlated pairs of cavity photons. The photon pairs in the ground
state are virtual and cannot escape the cavity if its parameters
are time-independent\cite{Ciuti_PRA}.

In order to release these bound photons into extracavity
radiation, the quantum vacuum has to be modulated in time. The
recent experimental demonstration of a wide tunability of the
cavity parameters (in particular of the vacuum Rabi frequency
$\Omega_R$ \cite{Aji_APL}) via an external gate bias and the
possibility of ultrafast modulation \cite{SNS_unpub} makes the
present system a very promising one in view of the observation of
quantum vacuum radiation. A first theoretical study for an ideal
isolated cavity \cite{Ciuti_vacuum} has indeed suggested that a
non-adiabatic switch-off of $\Omega_R$ results in a significant
number of photon pairs being emitted from the cavity.

A general theory which includes the effect of non-radiative and
radiative losses for arbitrary time-modulations is however still
missing, as well as a complete characterization of the intensity
and spectral signatures of the emitted radiation for realistic
cavity systems. In this letter, we address these fundamental
issues by developing a theory of the quantum vacuum radiation for
an arbitrary modulation of the microcavity properties. The
properties of the extracavity emission for the most promising case
of a periodic modulation of $\Omega_R$ are calculated by means of
the generalized input-output formalism \cite{Ciuti_PRA}:
remarkably, the emitted quantum vacuum radiation turns out to be
much stronger than the radiation by spurious effects such as black
body emission. The instability regions in which the vacuum
modulation produces a parametric oscillation of the cavity field
are identified and shown to be within experimental reach.

\begin{figure}[t!]
\begin{center}
\includegraphics[width=8cm]{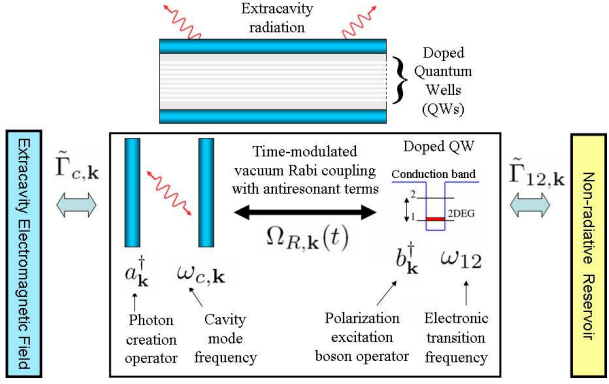}
\caption{\label{sketch} (color online). Top: a sketch of the
considered semiconductor planar microcavity system. Bottom: a
scheme of the quantum model. }
\end{center}
\end{figure}

A theoretical description of the system can be obtained by means
of the formalism developed in~\cite{Ciuti_vacuum,Ciuti_PRA}. The
photon mode in the planar microcavity and the bright intersubband
electronic excitation of the doped quantum well system (see Fig.
\ref{sketch}) are described as two bosonic modes. Given the
translational symmetry of the system along the cavity plane, the
in-plane wavevector $\kk$ is a good quantum number. The creation
operators for respectively a cavity photon and an electronic
excitation of wavevector $\kk$ are denoted by $\ahd_\kk$ and
$\bhd_\kk$. The in-plane dispersion relation of the cavity-photon
is defined as $\omega_{c,\kk}$, while the frequency $\omega_{12}$
of the intersubband excitation is taken dispersionless. As
explained in detail in Ref. \cite{Ciuti_vacuum}, the
electric-dipole coupling between one cavity photon and one bright
intersubband excitation is quantified by the vacuum Rabi frequency
$\Omega_{R,k} = \left ( \frac{2 \pi e^2}{\epsilon_{\infty}m _0
L^{\rm eff}_{\rm cav}} ~\sigma_{el} ~N^{eff}_{QW}~f_{12} \sin^2
\theta \right  )^{1/2}~$, where $L^{\rm eff}_{\rm cav}$ is the
effective length of the cavity mode, $\epsilon_{\infty}$ the
dielectric constant of the cavity spacer, $\sigma_{el}$ the
density of the two-dimensional electron gas, $N^{eff}_{QW}$ the
effective number of quantum wells, $f_{12}$ the oscillator
strength of the intersubband transition and $\theta$ is the
intracavity photon propagation angle such that $\sin\theta=c k
/(\omega_{12} \sqrt{\epsilon_{\infty}})$.
The second quantization Hamiltonian of the present cavity system
reads \begin{equation}
H=\frac{1}{2}\,\vhd_{\kk}\,\eta\,\Mc_\kk\,\vh_{\kk} \label{Hamilt}
\end{equation}
where the column vector of operators $\vh_{\kk}$ is defined
as \mbox{
$\vh_{\kk}\equiv
(
\ah_{\kk},\bh_{\kk},\ahd_{-\kk},\bhd_{-\kk}
)^T$,}
$\eta$ is the diagonal metric $\eta=\textrm{diag}[1,1,-1,-1]$ and
the Hopfield-Bogoliubov matrix $\Mc_\kk$ is defined as
\begin{equation}
\Mc_\kk\equiv \left(
\begin{array}{cccc}
\omega_{c,\kk}+2D_{\kk} & i \Omega_{R,\kk}&
2D_{\kk}&-i\Omega_{R,\kk}\\
-i\Omega_{R,\kk}&\omega_{12} &-i\Omega_{R,\kk}&0
\\
-2 D_{\kk}&-i\Omega_{R,\kk}
& -\omega_{c,\kk}-2D_{\kk}&i\Omega_{R,\kk}\\
-i\Omega_{R,\kk}&0&-i\Omega_{R,\kk}&
-\omega_{12,\kk}
\\
\end{array} \right ).
\nonumber
\end{equation}
For a quantum well, the diamagnetic coupling constant is
approximately $D_{\kk}\simeq
\Omega_{R,\kk}^2/\omega_{12}$~\cite{Ciuti_vacuum}. The
ultra-strong coupling regime is characterized by a value of
$\Omega_{R,\kk}$ comparable to $\omega_{c,\kk}$ and $\omega_{12}$.
In this regime, a central role is played by the anti-resonant
light-matter coupling terms corresponding to the off-diagonal
(1,3), (1,4), (2,3), (2,4) terms of $\Mc$ (and their transposed).
In the following, a general time-dependence of the Rabi frequency
is considered:
$\Omega_{R,\kk}(t)=\bar{\Omega}_{R,\kk}+\Omega^{mod}_{R,\kk}(t)$
and $D_k(t) = \bar{D}_k + D^{mod}_{k}(t)$.

Non-radiative as well as radiative losses will be taken into
account by means of the generalized input-output formalism
developed in~\cite{Ciuti_PRA}: the system is in interaction with
two baths of harmonic oscillators, which are responsible for
dissipation and fluctuations of both the cavity-photon and the
electronic polarization fields. The radiative and non-radiative
complex damping rates are denoted by ${\tilde
\Gamma}_{c,\kk}(\omega)$ and ${\tilde \Gamma}_{12,\kk}(\omega)$.
The real part (zero for $\omega < 0$~\cite{Ciuti_PRA}) quantifies
the frequency-dependent losses, while the imaginary part
represents the Lamb-shift of the mode due to the coupling to the
external bath. The resulting quantum Langevin equations are
conveniently written in frequency space as the vector equation:
\begin{equation}
\label{langevin} \int_{-\infty}^{+\infty}\!\!\!\! d\omega'\,
\left[\bar\Mc_{\kk,\omega} \,\delta(\omega-\omega')+
\Mc^{mod}_{\kk,\omega-\omega'}\right] \,\tilde{v}_{\kk}(\omega')=
-i\tilde{\mathcal F}_{\kk}
\end{equation}
where $\tilde{v}_\kk(\omega)$ is the Fourier transform of the
operator vector $\vh_\kk(t)$ and the quantum Langevin operator
vector
\begin{equation}
\tilde{\mathcal F}_{\kk}=\left( \tilde{F}_{c,\kk}(\omega),
\tilde{F}_{12,\kk}(\omega), \tilde{F}^\dagger_{c,-\kk}(-\omega),
\tilde{F}^\dagger_{12,-\kk}(-\omega) \right)^T \nonumber
\end{equation}
takes into account the quantum fluctuations due to the coupling to the baths.
The matrix
\begin{multline}
\bar\Mc_{\kk,\omega}={\mathcal
  M}_\kk-\mathbf{1}\,\omega
\\
-i\,\textrm{diag}[ \tilde{\Gamma}_{c,\kk}(\omega),
\tilde{\Gamma}_{12,\kk}(\omega),
\tilde{\Gamma}^*_{c,-\kk}(-\omega),
\tilde{\Gamma}^*_{12,-\kk}(-\omega) ] \nonumber
\end{multline}
summarizes the time-independent properties of the system, while
${\Mc}^{mod}_{\kk,\omega}$ describes the time-modulation. For the
case of a time-dependent vacuum Rabi frequency
$\Omega_{R,\kk}(t)$, this has the form:
\begin{equation}
{\Mc}^{mod}_{\kk,\omega}=\left(
\begin{array}{cccc}
2\tilde{D}^{mod}_{\kk,\omega}&i\tilde{\Omega}^{mod}_{R,\kk,\omega} &2\tilde{D}^{mod}_{\kk,\omega}&-i\tilde{\Omega}^{mod}_{R,\kk,\omega}\\
-i\tilde{\Omega}^{mod}_{R,\kk,\omega}&0&-i\tilde{\Omega}^{mod}_{R,\kk,\omega}&0\\
-2\tilde{D}^{mod}_{\kk,\omega}&-i\tilde{\Omega}^{mod}_{R,\kk,\omega}&-2\tilde{D}^{mod}_{\kk,\omega}&i\tilde{\Omega}^{mod}_{R,\kk,\omega}\\
-i\tilde{\Omega}^{mod}_{R,\kk,\omega}&0&-i\tilde{\Omega}^{mod}_{R,\kk,\omega}&0
\end{array}
\right)~, \nonumber
\end{equation}
where ${\tilde \Omega}^{mod}_{R,\kk,\omega}$ and
$\tilde{D}_{\kk,\omega}^{mod}$ are the Fourier transforms of
respectively $\Omega^{mod}_{R,\kk}(t)$ and $D_\kk^{mod}(t)$.

The exact solution of (\ref{langevin}) is given by
\begin{equation}
 \label{partial1}
\tilde{v}_{\kk}(\omega) = -i \int_{-\infty}^{\infty} d\omega'
{\Gc}_{\kk}(\omega,\omega') \tilde{\mathcal F}_{\kk}(\omega')~,
\end{equation}where $\Gc_{\kk}(\omega,\omega')$ is the
inverse of ${\Mc}_{\kk}(\omega,\omega')\equiv
\bar\Mc_{\kk,\omega'}
\delta(\omega-\omega')+{\Mc}^{mod}_{\kk,\omega-\omega'}$, i.e.,
\begin{equation}
\label{inverse} \int_{-\infty}^{\infty} d\omega' \sum_{s}
\Gc^{rs}_{\kk}(\omega,\omega'){\Mc}_{\kk}^{st}(\omega',\omega'')\equiv
\delta_{rt}\delta(\omega-\omega''). \nonumber
\end{equation}
Using the input-output scheme~\cite{Ciuti_PRA}, we obtain the
spectral density of emitted photons from the cavity
$S_\kk^{out}(\omega)$ as a function of the incident one
$S_\kk^{in}(\omega)$ and the quantum Langevin forces
$\tilde{\mathcal F}_{\kk}$:
\begin{widetext}
\begin{multline}
S_{\kk}^{out}(\omega)= S_{\kk}^{in}(\omega)
-\frac{i}{2\pi}\int_{-\infty}^{\infty} d\omega' \sum_{s}
{\Gc}_{\kk}^{*1s}(\omega,\omega')\tilde{\mathcal F}_{\kk}^{\dagger
\, s}(\omega') \tilde{\mathcal F}_{\kk}^1(\omega)
+\frac{i}{2\pi}\int_{-\infty}^{\infty} d\omega' \sum_{s}
{\Gc}^{1s}_{\kk}(\omega,\omega') \tilde{\mathcal F}^{\dagger
\,1}_{\kk}(\omega)\tilde{\mathcal F}_{\kk}^s(\omega) \\
+\frac{1}{\pi}\Re(\tilde{\Gamma}_{c,\kk}(\omega))
\int_{-\infty}^{\infty}\!\!\!
d\omega'\,\int_{-\infty}^{\infty}\!\!\! d\omega'' \,\sum_{rs}
\bar{\Gc}_{\kk}^{*1r}(\omega,\omega') {\Gc}_{\kk}^{1s}
(\omega,\omega'')\tilde{\mathcal F}_{\kk}^{\dagger \,
r}(\omega')\tilde{\mathcal F}_{\kk}^s(\omega'')~. \label{partial2}
\end{multline}
\end{widetext}
If one is interested in the quantum vacuum radiation due to the
time-modulation of the cavity parameters, a vacuum state has to be
considered for the input state, so that $\langle
S_\kk^{in}(\omega)\rangle=0$, and the fluctuating quantum Langevin
forces acting on the cavity-photon and the electronic polarization
modes ($j,j'\in\{c,12\}$) are such that:
\begin{eqnarray}
\left\langle\tilde{F}_{j,\kk}(\omega)\,\tilde{F}^\dagger_{j',\kk'}(\omega')\right\rangle
&=&4\pi\,\delta(\omega-\omega')\,\delta_{jj'}\,
\textrm{Re}[\Gamma_{j,\kk}(\omega)]
\,\delta_{\kk,\kk'} \nonumber \\
\left\langle\tilde{F}^\dagger_{j,\kk}(\omega)\,\tilde{F}_{j',\kk'}(\omega')\right\rangle
&=&\left\langle\tilde{F}_{j,\kk}(\omega)\,\tilde{F}_{j',\kk'}(\omega')\right\rangle=0.
\end{eqnarray}
This implies that only the last term of (\ref{partial2}) gives a
finite contribution to the emitted radiation. After some algebra,
we get :
\begin{multline}
\langle S_{\kk}^{out}(\omega) \rangle =
4\Re(\tilde{\Gamma}_{c,\kk}(\omega)) \int_{0}^{\infty}
\!\!\!\!\!d\omega'\, \left \{ \lvert
\bar{\Gc}_{\kk}^{13}(\omega,-\omega')\rvert^2
\Re[\tilde{\Gamma}_{c,\kk}(\omega')] \right .
\\ \left .
+\lvert \bar{\Gc}_{\kk}^{14}(\omega,-\omega')\rvert^2
\Re[\tilde{\Gamma}_{12,\kk}(\omega')] \right \}~.\label{Sout}
\end{multline}
The total number of emitted photons with in-plane wave-vector
$\kk$ is given by $N^{out}_{\kk}=
\int_{-\infty}^{\infty} \langle S_{\kk}^{out}(\omega)
\rangle d\omega$.
Note that in the absence of anti-resonant couplings in the Hamiltonian
(\ref{Hamilt}), $\bar{\Gc}_{\kk}^{13}=\bar{\Gc}_{\kk}^{14}=0$ giving a
vanishing emitted radiation.

This general theory can be applied to calculate the intensity of
quantum vacuum radiation emitted by the cavity for an arbitrary
modulation of the cavity parameters and for arbitrary
frequency-dependent losses. In the following, we shall focus
ourselves on the case of a periodic modulation of the vacuum Rabi
frequency, i.e.
\begin{equation}
\Omega^{mod}_{R,\kk}(t)=\Delta\Omega_{R,\kk}\cos(\omega_{mod}t).
\label{Omod}
\end{equation}
If the modulation frequency is tuned on resonance with the cavity
modes, one expects~\cite{Reynaud} that the quantum vacuum
radiation can be strongly enhanced as compared to the case of a
single sudden change of $\Omega_{R,\kk}$ discussed
in~\cite{Ciuti_vacuum}. In the stationary state, the relevant
quantity to characterize the intensity of the emission is the
total number of photons emitted per unit time
$dN^{out}_{\mathbf{k}}/dt$.
\begin{figure}[htbp]
\begin{center}
\includegraphics[width=8cm]{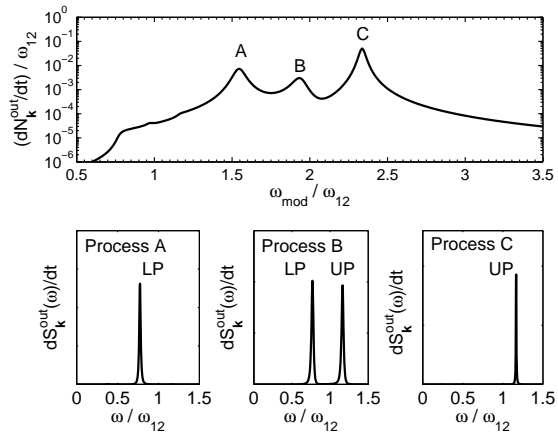}
\caption{\label{spectra} Top panel: rate of emitted photons
$dN_{\kk}^{out}/dt$ (in units of $\omega_{12}$) as a function of
the normalized modulation frequency $\omega_{mod}/\omega_{12}$.
Parameters: $(\omega_{c,\kk}+ 2D_k)/\omega_{12} = 1$,
$\Gamma/\omega_{12} = 0.025$, $\bar{\Omega}_{R,\kk}/\omega_{12} =
0.2$, $\Delta \Omega_{R,\kk}/\omega_{12} = 0.04$. Note that, due
to the scaling properties of the present model, the results do not
depend on the specific value of $\omega_{12}$. The letters A,B,C
indicate three different resonantly enhanced processes. Bottom
panels: the spectral density (arb. u.) for the processes A, B, C
respectively. The resonant peaks occur at the LP (Lower Polariton)
and/or UP (Upper Polariton) frequency.}
\end{center}
\end{figure}
\begin{figure}[htbp]
\begin{center}
\includegraphics[width=8cm]{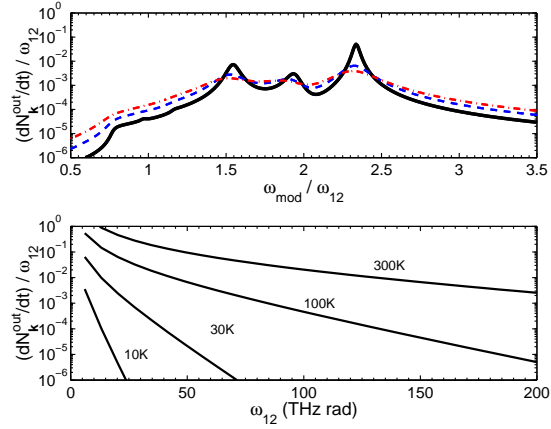}
\caption{(color online). \label{damping}Top panel: rate
$dN^{out}_{\mathbf{k}}/dt$ (in units of $\omega_{12}$) of emitted
photons as a function
  of   $\omega_{mod}/\omega_{12}$ for different values of the damping
$\Gamma/\omega_{12}=0.025$ (solid), $0.05$ (dashed), $0.075$
(dot-dashed). Other parameters as in Fig. \ref{spectra}. Bottom
panel: normalized rate of emitted photons from a black-body
emitter as a function of $\omega_{12}$ for different
temperatures.}
\end{center}
\end{figure}

Predictions for the rate $dN^{out}_{\mathbf{k}}/dt$ (in units of
$\omega_{12}$) of emitted photons as a function of the modulation
frequency $\omega_{mod}$ are shown in the top panel of
Fig.\ref{spectra}. For the sake of simplicity, a
frequency-independent damping rate has been considered $\Re
\{\tilde{\Gamma}_{c,\kk}(\omega > 0) \}= \Re \{
\tilde{\Gamma}_{12,\kk}(\omega > 0) \} = \Gamma$; the imaginary
part has been consistently determined via the Kramers-Kronig
relations~\cite{Ciuti_PRA}. Values inspired from recent
experiments \cite{Dini_PRL,Aji_APL,SNS_unpub} have been used for
the cavity parameters. The structures in the integrated spectrum
shown in the top panel of Fig.\ref{spectra} can be identified as
resonance peaks when the modulation is phase-matched. Indeed, the
vacuum modulation induces the creation of pairs of real cavity
polaritons. This sort of nonlinear parametric process is enhanced
when the phase-matching condition
$r\,\omega_{mod}=\omega_{j,\kk}+\omega_{j',-\kk}$ is fulfilled,
$r$ being a generic positive integer number, and
$j,j'\in\{LP,UP\}$. The dominant features A,B,C are the three
lowest-order $r=1$ peaks corresponding to the processes where
either two Lower Polaritons (LPs), or one LP and one Upper
Polariton (UP), or two UP's are generated. This interpretation is
supported by the spectral densities plotted in the three lower
panels of Fig.\ref{spectra} for modulation frequencies
corresponding to respectively A,B,C peaks. In each case, the
emission is strongly peaked at the frequencies of the final
polariton states involved in the process; for the parameter
chosen, we have indeed~\cite{Ciuti_vacuum,Ciuti_PRA}
$\omega_{LP,\kk} \simeq \omega_{12} - \Bar{\Omega}_{R,\kk} = 0.8\,
\omega_{12}$ for the lower polariton and $\omega_{UP,k} \simeq
\omega_{12} + \Bar{\Omega}_{R,\kk} = 1.2\, \omega_{12}$ for the
upper polariton. The shoulder and the smaller peak at
$\omega_{mod}/\omega_{12}<1$ can be attributed to $r=2$ processes,
while higher order processes require a weaker damping to be
visible.

More insight into the properties of the quantum vacuum emission
are given in Fig.\ref{damping}. In the top panel, the robustness
of the emission has been verified for increasing values of the
damping rate $\Gamma$, the resonant enhancement is quenched, but
the main features remain unaffected even for rather large damping
rates. In the bottom panel, comparison with the black body
emission in the absence of any modulation is made: the total rate
of emitted black body photons at given $\kk$ is shown as a
function of $\omega_{12}$ (ranging from the terahertz to the mid
infrared range) for $\kk$ correspoding to an intracavity photon
propagation angle of $60^{\circ}$ and different temperatures. Note
how the black body emission decreases almost exponentially with
$\omega_{12}$, while the quantum vacuum radiation, being a
function of $\bar{\Omega}_{R,\kk}/\omega_{12}$ only, linearly
increases with $\omega_{12}$ at fixed
$\bar{\Omega}_{R,\kk}/\omega_{12}$. From this plot, one is
quantitatively reassured that for reasonably low temperatures the
quantum vacuum radiation can exceed the black-body emission even
by several orders of magnitude.
\begin{figure}[htbp]
\begin{center}
\includegraphics[width=8cm]{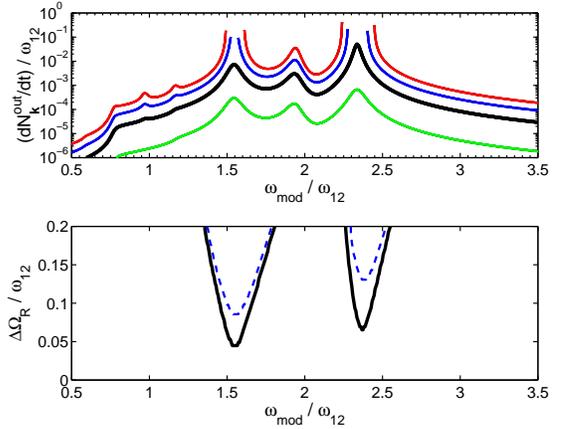}
\caption{(color online). \label{instability}Top panel: rate
$dN^{out}_{\mathbf{k}}/dt$ (in units of $\omega_{12}$) of emitted
photons as a function of $\omega_{mod}/\omega_{12}$ for different
values of
    the normalized modulation amplitude $\Delta
    \Omega_{R,k}/\omega_{12}=0.01, 0.04, 0.07, 0.1$ (from bottom to
    top). Other parameters as in Fig. \ref{spectra}. Bottom panel:
    instability boundaries for $\Gamma/\omega_{12} =
0.025$ (solid), $0.05$ (dashed). Above the lines, the system is
    parametrically unstable. }
\end{center}
\end{figure}
The increase of the emitted intensity versus the modulation
amplitude $\Delta\Omega_{R,\kk}/\omega_{12}$ is shown in the top
panel of Fig.\ref{instability}. In particular, note the strongly
superlinear increase of the emission intensity around the A and C
resonance peaks. In these regions, if the modulation amplitude is
large enough, the system can even develop a parametric
instability, the incoherent quantum vacuum radiation being
replaced by a coherent parametric oscillation~\cite{OPO}. Above
the instability threshold, the results obtained from the solution
of Eq. (\ref{langevin}) in Fourier space are no longer valid,
being the field amplitudes exponentially growing with time. Hence
they are not shown here. The instability boundaries for parametric
oscillation can be calculated applying the Floquet
method\cite{Floquet} to the mean-field equations for the
intracavity fields $\langle a_{\kk} \rangle$ and $\langle b_{\kk}
\rangle$. The result is shown in the bottom panel of
Fig.\ref{instability} as a function of $\omega_{mod}/\omega_{12}$
and $\Delta \Omega_{R,k}/\omega_{12}$: agreement with the position
of the vertical asymptotes of the spectra in the top panel of of
Fig.\ref{instability} is found.


In conclusion, we have presented a complete theory with exact
solutions for the quantum vacuum radiation from a semiconductor
microcavity with a time-modulated vacuum Rabi frequency. In order
to isolate it from spurious effects such as black-body radiation,
the main signatures of the quantum vacuum radiation as a function
of the modulation parameters have been characterized. Our results
show that semiconductor microcavities in the ultrastrong coupling
regime are a very promising system for the observation of quantum
vacuum radiation phenomena.


\begin{thebibliography}{}

\bibitem{RMP} M. Kardar and R. Golestanian, Rev. Mod. Phys.
{\bf 71}, 1233 (1999).
\bibitem{dyn_Casimir} G. T. Moore, J. Math. Phys. (N.Y.) {\bf 11}, 2679
  (1970); S. A. Fulling and P. C. W. Davies, Proc. R. Soc. London A
  {\bf 348}, 393 (1976)
\bibitem{Yablonovitch} E. Yablonovitch, Phys. Rev. Lett. ${\bf 62}$, 1742 (1989).
\bibitem{Hawking} W. G. Unruh, Phys. Rev. D {\bf 10}, 3194 (1974);
  S. W. Hawking, Nature (London) {\bf 248}, 30 (1974).
\bibitem{Reynaud} A. Lambrecht, M. T. Jaekel, and S. Reynaud,
Phys. Rev. Lett. {\bf 77}, 615 (1996). For a recent review, see:
A. Lambrecht, J. Opt. B: Quantum Semiclass. Opt. {\bf 7}, S3 (2005).
\bibitem{epsilon_var} V. V. Dodonov, A. B. Klimov, D. E. Nikonov,
  Phys. Rev. A {\bf 47}, 4422 (1993); C. K. Law, Phys. Rev. A {\bf
  49}, 433 (1994).
\bibitem{Braggio} C. Braggio, {\em et al.},
  Rev. Sc. Instr. {\bf 75}, 4967 (2004);
  Europhys. Lett. {\bf 70} 754 (2005).
\bibitem{Dini_PRL} D. Dini, R. Kohler, A. Tredicucci, G. Biasiol, and L. Sorba,
Phys. Rev. Lett. {\bf 90}, 116401 (2003).
\bibitem{Dupont} E. Dupont, H. C. Liu, A. J. SpringThorpe, W. Lai, and M. Extavour, Phys. Rev.
B {\bf 68}, 245320 (2003).
\bibitem{Aji_APL} A. A. Anappara, A. Tredicucci, G. Biasiol, L. Sorba, Appl.
Phys. Lett. {\bf 87}, 051105 (2005).
\bibitem{SST} R. Colombelli, C. Ciuti, Y. Chassagneux, C.
Sirtori, Semicond. Sci. Technol. {\bf 20}, 985 (2005).
\bibitem{SNS_unpub} A. A. Anappara, A. Tredicucci, F. Beltram, G. Biasol, L. Sorba,
Appl. Phys. Lett. {\bf 89}, 171109 (2006).
\bibitem{Luca} L. Sapienza, A. Vasanelli, C. Sirtori {\it et al.}, in preparation
\bibitem{Ciuti_vacuum} C. Ciuti, G. Bastard, I. Carusotto, Phys.
Rev. B {\bf 72}, 115303 (2005).
\bibitem{Ciuti_PRA} C. Ciuti, I. Carusotto, Phys. Rev. A {\bf 74},
  033811 (2006).
\bibitem{OPO} D. F. Walls and G. J. Milburn, {\em Quantum
  Optics} (Springer, Berlin, 1994).
\bibitem{Floquet} R. Grimshaw, {\em Nonlinear Ordinary Differential equations} (CRC Press,
1993).

\end{thebibliography}
\end{document}